# Existence and Uniqueness of the Solution to a Nonlinear Differential Equation with Caputo Fractional Derivative in the Space of Continuously Differentiable Functions

Sunae Pak and Myongha Kim

Department of Mathematics, **Kim Il Sung** University, D.P.R.K
Email: myongha_kim@yahoo.com

*Abstract*: In the paper, we considered the existence and uniqueness of the global solution in the space of continuously differentiable functions for nonlinear differential equation with the Caputo fractional derivative of general form. Here we considered the existence and uniqueness of the solution of initial value problem:

$${}^c D_{0+}^{\alpha} y(t) = f(t, {}^c D_{0+}^{\alpha_1} y(t), \cdots, {}^c D_{0+}^{\alpha_m} y(t)), \quad D^j y(t)|_{t=+0} = b_j \in R, \quad j = 0, 1, \cdots, n-1.$$

## 1. Introduction

In recent years there have been considerable interests in the theory and applications of fractional differential equations ([1, 5, 6, 7]). Indeed, it is well known that fractional derivative is excellent tool for description of memory and heredity effects. (See [7].) Many authors studied linear fractional equations and good described material on them are found in [1, 5, 6].

On the other hand, recent studies on fractional calculus are focused on nonlinear equations. (See [2, 3, 4, 7].)

In [2], the authors considered the existence and uniqueness of the local solution to fractional differential equations with the Riemann-Liouville fractional derivative to in the case of monomial. In [3], the authors considered the existence and uniqueness of the global solution to the fractional differential equations with the Riemann-Liouville fractional derivative of general form in the space of summable functions.

In [4, 5] the authors considered the existence and uniqueness of the global solution to the nonlinear differential equations with the Caputo fractional derivative in the case of monomial.

In this paper, we consider the existence and uniqueness of the global solution to the nonlinear fractional differential equations with the Caputo fractional derivative in the space of continuously differentiable functions.

## 2. Preliminaries

**Definition 1.**[1, 5] Let's denote by $R = (-\infty, +\infty)$, $R_+ = (0, +\infty)$. Let $\Omega = [a,b] (-\infty \leq a < b \leq +\infty)$ and $m \in N_0 = \{0,1,\cdots,n-1\}$.

We denote by $C^m(\Omega)$ the space of functions $f$ which are $m$ times continuously differentiable on $\Omega$ with norm

$$\|f\|_{C^m} = \sum_{k=0}^{m} \|f^{(k)}\|_C = \sum_{k=0}^{m} \max_{x \in \Omega} |f^{(k)}|, \; m \in N_0.$$

In particular, $C^0(\Omega) \equiv C(\Omega)$ is the space of continuous functions $f$ on $\Omega$ with the norm

$$\|f\|_C = \max |f^{(k)}|.$$

$C_\gamma[a,b]$ of such functions $f$ that $y(t) \in C_\gamma^{\alpha,n-1}[0,T]$ and $f$ itself is defined on $(a,b]$ and define the norm of this space as follows:

$$\|f\|_{C_\gamma} = \|(x-a)^\gamma f(x)\|_C, \; C_0[a,b] = C[a,b].$$

For $n \in N$ we denote by $C_\gamma^n[a,b]$ the space of functions $f$ which are $n-1$ times continuously differentiable on $[a,b]$ and have the derivative $f^{(n)}(x)$ such that $f^{(n)}(x) \in C_\gamma[a,b]$. We define the norm of this space as follows:

$$\|f\|_{C_\gamma^n} = \sum_{k=0}^{n-1}\|f^{(k)}\|_C + \|f^{(n)}\|_{C_\gamma}$$

And we denote as follows:

$$C_\gamma^0[a,b] = C_\gamma[a,b].$$

**Definition 2.** [1, 5]  Let $\alpha \in R_+, f \in C_\gamma[0.T], 0 \leq \gamma < 1$. The *fractional integrals* $I_{0+}^\alpha f$ of order $\alpha \, (\alpha > 0)$ of function $f$ in the means of *Riemann-Liouville* are defined by

$$I_{0+}^\alpha f(t) = \frac{1}{\Gamma(\alpha)}\int_0^t (t-\tau)^{\alpha-1} f(\tau)d\tau, t > 0.$$

In particular, we denote by

$$I^0 f(t) = f(t).$$

**Definition 3.** [1, 5]  Let $n \in N$, $n-1 < \alpha \leq n$, and $0 \leq \gamma < 1$. When $I^{n-\alpha} f \in C_\gamma^n[0, T]$, the *Riemann-Liouville fractional derivatives* $D_{a+}^\alpha f$ of order $\alpha \in R_+ \, (\alpha > 0)$ are defined by

$$D_{0+}^\alpha f(t) = D^n I_{0+}^{n-\alpha} f(t),$$

where

$$D^n = \frac{d^n}{dt^n}.$$

**Definition 4.** [5]  When $n \in N$, $n-1 < \alpha \leq n$, the *Caputo fractional derivatives* ${}^c D_{0+}^\alpha f$ of order $\alpha \, (\alpha > 0)$ of function $f$ are defined by

$${}^c D_{0+}^\alpha f(t) = D_{0+}^\alpha (f(t) - \sum_{k=0}^{n-1} \frac{f^{(k)}(0)}{k!} t^k).$$

**Lemma 1.** [5] (Banach fixed point theorem)  *Let $(U, d)$ be a nonempty complete metric space, let $0 \leq \omega < 1$, and let $T : U \to U$ be the map such that for every $u, v \in U$ the relation*

$$d(Tu, Tv) \leq \omega d(u,v), 0 \leq \omega < 1$$

holds. Then, the operator $T$ has a unique fixed point $u^* \in U$. Furthermore, if $T^K (K \in N)$ is the sequence of operators defined by

$$T^1 = T, \text{ and } T^K = TT^{K-1}, (K \in N \setminus \{1\}),$$

then for any $u_0 \in U$, the sequence $\{T^K u_0\}\big|_{k=1}^{\infty}$ converges to the above fixed point $u^*$.

**Lemma 2.**[5] Let $\alpha > 0$, $0 \leq \gamma < \alpha - n + 1$ and $f \in C_\gamma[0,T]$. then holds:

$$\lim_{t \to +\infty} I^\alpha f(t) = 0, \quad I^\alpha f(t) \in C[0, T].$$

## 3. Main Theorems

Let's consider the following nonlinear fractional differential equations with the Caputo fractional derivative:

$$^c D_{0+}^\alpha y(t) = f(t, {}^c D_{0+}^{\alpha_1} y(t), \cdots {}^c D_{0+}^{\alpha_m} y(t)). \tag{1}$$

where

$$\alpha, \alpha_h \in R_+, \quad h = 1, \cdots, m \text{ and } \alpha > 0, \alpha > \alpha_1 > \cdots > \alpha_m \geq 0.$$

Let's $n, n_h$ is natural number such that $n - 1 < \alpha \leq n$, $n_h - 1 < \alpha_h \leq n_h$, $h = 1, \cdots m$. And initial condition to the equation (1) is followings:

$$D^j y(t)\big|_{t=+0} = b_j \in R, \quad j = 0, 1, \cdots, n-1. \tag{2}$$

Volterra integral equations corresponding to the initial value problem (1), (2) is as follows:

$$y(t) = \sum_{j=0}^{n-1} b_j \Phi_{j+1}(t) + I_{0+}^\alpha f(t, {}^c D_{0+}^{\alpha_1} y(t), \cdots, {}^c D_{0+}^{\alpha_m} y(t)) \tag{3}$$

If $\alpha = n \in N$, then the equation (1) is as follows:

$$y^{(n)} = f(t, {}^c D_{0+}^{\alpha_1} y(t), \cdots, {}^c D_{0+}^{\alpha_m} y(t)). \tag{4}$$

The integral equation corresponding to the initial value problem (4) and (2) is as follows:

$$y(t) = \sum_{j=0}^{n-1} b_j \Phi_{j+1}(t) + I_{0+}^n f(t, {}^c D_{0+}^{\alpha_1} y(t), \cdots, {}^c D_{0+}^{\alpha_m} y(t)) \qquad (5)$$

**Theorem 1.** *Let's* $\alpha > 0$.

① Let $\alpha = n \in N$ and $f : [0,T] \times R^m \to R$ be a function such that $f(t, y) \in C[0,T]$ for any $y \in \mathrm{R}^m$. Then $y(t) \in C^n[0,T]$ satisfies the relations (4) and (2) if and only if $y(t) \in C^n[0,T]$ satisfies the integral equation (5).

② Let $n, n_1 \in N$, $n-1 < \alpha < n$, $n > n_1$, $0 \le \gamma < \alpha - n + 1$ and $f : [0,T] \times \mathrm{R}^m \to R$ be a function such that $f(t,y) \in C_\gamma[0,T]$ for any $y \in \mathrm{R}^m$. Then $y(t) \in C^{n-1}[0,T]$ satisfies the relations (1) and (2) if and only if $y(t) \in C^{n-1}[0,T]$ satisfies the integral equation (3).

**Proof.** We skip the proof of ① and prove the necessity of ②.

Let $y(t) \in C^n[0,T]$ satisfy the initial value problem (1) and (2). According to definition of Caputo and Riemann-Liouville fractional derivative

$$^c D_{0+}^\alpha y(t) = D_{0+}^\alpha (y(t) - \sum_{j=0}^{n-1} D^j y(0) \Phi_{j+1}(t)) = D^n I_{0+}^{n-\alpha} (y(t) - \sum_{j=0}^{n-1} D^j y(0) \Phi_{j+1}(t))$$

holds. By (1) and the condition on $f$, we have ${}^c D_{0+}^\alpha y(t) \in C_\gamma[0,T]$ and hence

$$D^n I_{0+}^{n-\alpha} \left( y(t) - \sum_{j=0}^{n-1} D^j y(0) \Phi_{j+1}(t) \right) \in C_\gamma[0, T].$$

Hence, by the definition of $C_\gamma^n[0,T]$, we have

$$I_{0+}^{n-\alpha} (y(t) - \sum_{j=0}^{n-1} D^j y(0) \Phi_{j+1}(t)) \in C_\gamma[0,T].$$

Clearly $C_\gamma^n[0, T] \subset AC^n[0, T]$ and if we let

$$g(t) := y(t) - \sum_{j=0}^{n-1} D^j y(0) \Phi_{j+1}(t),$$

then

$$g_{n-\alpha}(t) = I_{0+}^{n-\alpha} g(t) \in AC^n[0,T].$$

By (2.1.33) of [5], the following holds:

$$I_{0+}^{\alpha}\,{}^c D_{0+}^{\alpha} y(t) = I_{0+}^{\alpha} D_{0+}^{\alpha} [y(t) - \sum_{j=0}^{n-1} D^j y(0) \Phi_{j+1}(t)) =$$

$$= I_{0+}^{\alpha} D_{0+}^{\alpha} g(t) = g(t) - \sum_{k=1}^{n} \frac{g_{n-\alpha}^{(n-k)}(0)}{\Gamma(\alpha-k+1)} t^{\alpha-k}$$

Then the following holds:

$$Dg_{n-\alpha}(t) = D(g*\Phi_{n-\alpha})(t) = (Dg*\Phi_{n-\alpha})(t) + g(0)\Phi_{n-\alpha}(t) = (Dg*\Phi_{n-\alpha})(t),$$

because $g_{n-\alpha}(t) = I_{0+}^{n-\alpha} g(t) = (g*\Phi_{n-\alpha})(t)$ and $g(0)=0$.

On the other hand, $Dg(t) = Dy(t) - \sum_{j=1}^{n-1} D^j y(0) \Phi_j(t)$ and hence we get

$$Dg_{n-\alpha}(t) = I_{0+}^{n-\alpha}[Dy(t) - \sum_{j=1}^{n-1} \frac{y^{(j)}(0)}{(j-1)!} t^{j-1}].$$

By the same method we have

$$D^{n-k} g_{n-\alpha}(t) = (D^{n-k} g * \Phi_{n-\alpha})(t) = I_{0+}^{n-\alpha} D^{n-k} g(t)$$

$$= I_{0+}^{n-\alpha}[D^{n-k} y(t) - \sum_{j=n-k}^{n-1} \frac{y^{(j)}(0)}{(j-n+k)!} t^{j-n+k}]$$

$$= I_{0+}^{n-\alpha}[D^{n-k} y(t) - \sum_{j=n-k}^{n-1} \frac{y^{(j)}(0)}{(j-n+k)!} t^{j-n+k}].$$

Let

$$H(t) := D^{n-k} y(t) - \sum_{j=n-k}^{n-1} \frac{y^{(i)}(0)}{(j-n+k)!} t^{j-n+k}.$$

Then $H(t)$ is continuous function and $n-\alpha > 0$ and by lemma 2, we have

$$I_{0+}^{n-\alpha} H(0) = 0.$$

Therefore,

$$g_{n-\alpha}^{(n-k)}(0) = D^{n-k} g_{n-\alpha}(t)\big|_{t=+0} = I_{0+}^{n-\alpha} H(t)\big|_{t=+0} = 0. \qquad (6)$$

Apply the operator $I_{0+}^{\alpha}$ to both sides of equation (1) and consider (6) and (2),

then we obtain $y(t) - \sum_{j=0}^{n-1} b_j \Phi_{j+1}(t) = I_{0+}^{\alpha} f(t, {}^c D_{0+}^{\alpha_1} y(t), \cdots, {}^c D_{0+}^{\alpha_m} y(t))$. That is $y(t) \in C^{n-1}[0,T]$ satisfy the integral equation (3).

Now we prove the sufficiency. We'll prove the sufficiency of ②.

Let $y(t) \in C^{n-1}[0,T]$ satisfy the equation (3). That is

$$y(t) = \sum_{j=0}^{n-1} b_j \Phi_{j+1}(t) + I_{0+}^{\alpha} f(t, {}^c D_{0+}^{\alpha_1} y(t), \cdots {}^c D_{0+}^{\alpha_m} y(t)). \tag{7}$$

The right side of (7) is $n-1$ order differentiable, applying the operator $D^k$ to both sides of (7), we have for $k = 0, 1, \cdots, n-1$

$$D^k y(t) = \sum_{j=k}^{n-1} b_j \Phi_{j+1-k}(t) + I_{0+}^{\alpha-k} f(t, {}^c D_{0+}^{\alpha_1} y(t), \cdots, {}^c D_{0+}^{\alpha_m} y(t)). \tag{8}$$

Since

$$f(t, y) \in C_\gamma[0, T]$$

for any $y = ({}^c D_{0+}^{\alpha_1} y, \cdots, {}^c D_{0+}^{\alpha_m} y) \in R^m$ and $\alpha - k - \gamma > \alpha - n + 1 - \gamma > 0$, we have

$$I^{\alpha-k} f(t, {}^c D_{0+}^{\alpha_1} y, \cdots, {}^c D_{0+}^{\alpha_m} y) \in C[0, T], \tag{9}$$

$$\lim_{t \to +0} I^{\alpha-k} f(t, {}^c D_{0+}^{\alpha_1} y, \cdots, {}^c D_{0+}^{\alpha_m} y) = 0, \quad k = 0, 1, \cdots, n-1. \tag{10}$$

On the other hand, under consideration of the relation

$$\Phi_{j+1-k}(t)\Big|_{t=+0} = \begin{cases} 1, j = k \\ 0, j \neq k, \end{cases} \quad (j \geq k)$$

and relation (10) we let $t \to +0$ on the both sides of (8), then we obtain

$$D^k y(t)|_{t=+0} = b_k, \quad k = 0, 1, \cdots, n-1.$$

Hence the $y(t)$ satisfies relation (8) satisfy the initial condition (2). From (7) we have

$$y(t) - \sum_{j=0}^{n-1} b_j \Phi_{j+1}(t) = I_{0+}^{\alpha} f(t, {}^c D_{0+}^{\alpha_1} y(t), \cdots {}^c D_{0+}^{\alpha_m} y(t)).$$

Take fractional derivative of $\alpha$ order in means of Riemann-Liouville on both side of this relation and consider the conditions for $f$, then we obtained

$$D_{0+}^{\alpha}\left(y(t) - \sum_{j=0}^{n-1} b_j \Phi_{j+1}(t)\right) = D_{0+}^{\alpha} I_{0+}^{\alpha} f(t, {}^cD_{0+}^{\alpha_1} y(t), \cdots, {}^cD_{0+}^{\alpha_m} y(t)) =$$
$$= f(t, {}^cD_{0+}^{\alpha_1} y(t), \cdots, {}^cD_{0+}^{\alpha_m} y(t))$$

By definition 1.4, the left hand of the above expression is ${}^cD_{0+}^{\alpha} y(t)$ and thus

$$ {}^cD_{0+}^{\alpha} y(t) = f(t, {}^cD_{0+}^{\alpha_1} y(t), \cdots, {}^cD_{0+}^{\alpha_m} y(t)).$$

Hence $C^{n-1}[0,T] \ni y(t)$ denoted by (7) satisfies (1). That is $C^{n-1}[0, T] \ni y(t)$ satisfies the initial value problem (1),(2).(QED)

**Theorem 2**. *Let $T$ is arbitrary positive number. Let $\alpha = n \in N$, $n_i \in N$, $n_i - 1 < \alpha_i < n_i, i = 1, \cdots, m$. Let $f : [0,T] \times R^m \to R$ be a function and assume that $f(t, y) \in C[0, T]$ for every $(y_1, \cdots, y_m) \in R^m$ and $f$ satisfies the Lipschitzian condition on second variable $y$. Then there exists a unique solution of the initial value problem (4) and (2) in $C^n[0,T]$.*

**Theorem 3**. *Let $T$ is arbitrary positive number. Let $n, n_1 \in N$, $n - 1 < \alpha < n$, $n > n_1$, $0 \le \gamma < \alpha - n + 1$. Let $f : [0,T] \times Q \to R$ be a function and assume that $f(t, y) \in C_\gamma[0, T]$ for every $(y_1, \cdots, y_m) \in R^m$ and $f$ satisfies the Lipschitzian condition on second variable $y$. Then there exists a unique solution $y(t) \in C_\gamma^{\alpha, n-1}[0,T]$ of the initial value problem (1) and (2).*

The proofs of the theorem 2 and 3 can be done by the standard way of solving integral equations using Banach fixed point theorem(lemma 1) and here we omit them.